# NONRECIPROCAL TRANSVERSE PHOTONIC SPIN AND MAGNETIZATION-INDUCED ELECTROMAGNETIC SPIN-ORBIT COUPLING

Miguel Levy and Dolendra Karki
Physics Department, Michigan Technological University
Henes Center for Quantum Phenomena

ABSTRACT

A study of nonreciprocal transverse-spin angular-momentum-density shifts for evanescent waves in magneto-optic waveguide media is presented. Their functional relation to electromagnetic spin- and orbital-momenta is presented and analyzed. It is shown that the magneto-optic gyrotropy can be re-interpreted as the nonreciprocal electromagnetic spin-density shift per unit energy flux, thus providing an interesting alternative physical picture for the magneto-optic gyrotropy. The transverse spin-density shift is found to be thickness-dependent in slab optical waveguides. This dependence is traceable to the admixture of minority helicity components in the transverse spin angular momentum. It is also shown that the transverse spin is magnetically tunable. A formulation of electromagnetic spin-orbit coupling in magneto-optic media is presented, and an alternative source of spin-orbit coupling to non-paraxial optics vortices is proposed. It is shown that magnetization-induced electromagnetic spin-orbit coupling is possible, and that it leads to spin to orbital angular momentum conversion in magneto-optic media evanescent waves.

## INTRODUCTION

In 1939, F. J. Belinfante introduced a spin momentum density expression for vector fields to explain the spin of quantum particles and symmetrize the energy-momentum tensor [1]. For monochromatic electromagnetic waves in free-space, the corresponding time-averaged spin momentum density reads

$$\vec{p}_B = \frac{1}{2}\vec{\nabla}\times\vec{s}_B, \qquad (\text{Eq. 1})$$

and the spin angular momentum density is

$$\vec{s}_B = \text{Im}\frac{1}{2\omega}(\varepsilon_o \vec{E}^* \times \vec{E}). \qquad (\text{Eq. 2})$$

$\omega$ is the optical frequency and $\varepsilon_0$ the permittivity of free-space [2].

This spin angular momentum, in its transverse electromagnetic form, has merited much attention in recent years, as it can be studied in evanescent waves [3-7]. There are fundamental and practical reasons for this.

Until recently, the quantum field theory of the electromagnetic field has lacked a description of separate local conservation laws for the spin and orbital angular momentum-generating currents [7]. Whether such spin-generating momenta, as opposed to the actual spin angular momenta they induce, are indeed observable or merely 'virtual' is of fundamental interest. Moreover, if the electromagnetic spin and orbital momenta are separable, the question arises as to whether there are any photonic spin-orbit interaction effects. Bliokh and co-workers give a positive answer for non-paraxial fields. [7] Using the conservation laws proposed by these authors, we show that it is possible to magnetically induce electromagnetic spin-orbit coupling in magneto-optic media.

We know that the transverse optical spin is a physically meaningful quantity that can be transferred to material particles [3-8]. This has potentially appealing consequences for optical-tweezer particle manipulation, or to locate and track nanoparticles with a high degree of temporal and spatial resolution [9]. Thus, developing means of control for the transverse optical spin is of practical interest.

In this paper, we address the latter question for both spin momenta $\vec{p}_B$ and angular momenta $\vec{s}_B$. We show that their magnitudes and sense of circulation can be accessed and controlled in a single structure, and propose a specific configuration to this end. Explicit expressions for these physical quantities and for the spin-orbit coupling are presented. Moreover, we develop our treatment for nonreciprocal slab optical waveguides, resulting in a different response upon time reversals.

We consider the behavior of evanescent waves in transverse-magnetic (TM) modes in magnetic garnet claddings on silicon-on-insulator guides. This allows us to obtain explicit expressions for the transverse Belinfante spin momenta and angular momenta and to propose a means for magnetically controlling these objects, with potential applicability to nanoparticle manipulation.

MAGNETIC-GYROTROPY-DEPENDENT EVANESCENT WAVES

Consider a silicon-on-insulator slab waveguide with iron garnet top cladding, as in Fig. 1. The off-diagonal components of the garnet's dielectric permittivity tensor control the structure's magneto-optic response. Infrared 1550nm wavelength light propagates in the slab, in the presence of a magnetic field transverse to the direction of propagation.

The electromagnetic field-expressions in the top cladding for transverse magnetization (y-direction) and monochromatic light propagating in the z-direction are,

$$\vec{E} = \vec{E}_o(x,y)e^{i(\beta z - \omega t)}$$
$$\vec{H} = \vec{H}_o(x,y)e^{i(\beta z - \omega t)} \quad \text{(Eq. 3)}$$

Maxwell-Ampere's and Faraday's laws in ferrimagnetic media are

$$\vec{\nabla} \times \vec{H} = \varepsilon_o \hat{\varepsilon} \frac{\partial \vec{E}}{\partial t} = \varepsilon_o \begin{pmatrix} \varepsilon_c & 0 & ig \\ 0 & \varepsilon_c & 0 \\ -ig & 0 & \varepsilon_c \end{pmatrix} \frac{\partial \vec{E}}{\partial t} = -\varepsilon_o \begin{pmatrix} \varepsilon_c & 0 & ig \\ 0 & \varepsilon_c & 0 \\ -ig & 0 & \varepsilon_c \end{pmatrix} i\omega \vec{E} \quad \text{(Eq. 4)}$$

$$\vec{\nabla} \times \vec{E} = -\mu_o \frac{\partial \vec{H}}{\partial t} = \mu_o i\omega \vec{H} \quad \text{(Eq. 5)}$$

The off-diagonal component of the dielectric permittivity tensor $\hat{\varepsilon}$ is the gyrotropy parameter, parameterized by $g$.

We examine transverse-magnetic (TM) propagation in the slab. Vertical and transverse-horizontal directions are $x$, and $y$, respectively, $\beta$ is the propagation constant, and the wave equation in the iron garnet is given by

$$\frac{\partial^2}{\partial x^2}H_y + \left[k_o^2\left(\varepsilon_c - \frac{g^2}{\varepsilon_c}\right) - \beta^2\right]H_y = 0 \text{ , with } k_0 = 2\pi/\lambda \text{ , for wavelength } \lambda \ [10, 11]. \quad \text{(Eq. 6)}$$

Defining $\varepsilon_{eff} = \left(\varepsilon_c - \frac{g^2}{\varepsilon_c}\right)$ as an effective permittivity in the cover layer, we get:

$$H_y = H_c e^{-\gamma_{eff} x}, x > 0 \qquad \text{(Top cladding)} \qquad \text{(Eq. 7)}$$
$$H_y = H_f \cos(k_x x + \phi_c), -d < x < 0 \qquad \text{(Core)} \qquad \text{(Eq. 8)}$$
$$H_y = H_s \exp(\gamma_s(x+d)), x < -d \text{ ,} \qquad \text{(Substrate)} \qquad \text{(Eq. 9)}$$

where

$$\gamma_{eff} = \sqrt{\beta^2 - k_o^2 \varepsilon_{eff}} \text{ ,} \qquad \text{(Eq. 10)}$$
$$k_x = \sqrt{k_o^2 \varepsilon_f - \beta^2} \text{ ,} \qquad \text{(Eq. 11)}$$
$$\gamma_s = \sqrt{\beta^2 - k_o^2 \varepsilon_s} \qquad \text{(Eq. 12)}$$

$\varepsilon_f$, and $\varepsilon_s$ are the silicon-slab and substrate dielectric-permittivity constants, respectively, and $d$ is the slab thickness.

Solving for the electric-field components in the top cladding layer, we get,

$$E_z = i\frac{g\beta - \varepsilon_c \gamma_{eff}}{\omega \varepsilon_0 (\varepsilon_c^2 - g^2)} H_y$$
$$E_x = \frac{\beta \varepsilon_c - g\gamma_{eff}}{\omega \varepsilon_0 (\varepsilon_c^2 - g^2)} H_y \qquad \text{(Eq. 13)}$$

Notice that these two electric field components are $\pi/2$ out of phase, hence the polarization is elliptical in the cover layer, with optical spin transverse to the propagation direction. In addition, the polarization evinces opposite helicities for counter-propagating beams, as $E_z / E_x$ changes sign upon propagation direction reversal.

This result already contains an important difference with reciprocal non-gyrotropic formulations, where $E_z / E_x = -i\gamma/\beta$, and $\gamma$ the decay constant in the top cladding. Equation 13 depends on the gyrotropy parameter $g$, both explicitly and implicitly through $\beta$, and is therefore magnetically tunable, as we shall see below.

We emphasize that the magnitude and sign of the propagation constant $\beta$ change upon propagation direction reversal, and, separately, upon magnetization direction reversal. The difference between forward and backward propagation constants is also gyrotropy dependent. This nonreciprocal quality of magneto-optic waveguides is central to the proper functioning of certain on-chip devices, such as Mach-Zehnder-based optical isolators [10, 11].

As pointed out before, Eq. 2 applies to free-space Maxwell electromagnetism. In a dielectric medium, the momentum density expression must account for the electronic response to the optical wave. Minkowski's and Abraham's formulations describe the canonical and the kinetic electromagnetic momenta, respectively [12]. Here we will focus on Minkowski's version, $\vec{p} = \vec{D} \times \vec{B}$, as it is intimately linked to the generation of translations in the host medium, and hence to optical phase shifts, of interest in nonreciprocal phenomena. $\vec{D}$ is the displacement vector, and $\vec{B}$ the magnetic flux density.

Dual-symmetric versions of electromagnetic field theory in free space have been considered by various authors [2, 7, 8, 13]. However, the interaction of light and matter at the local level often has an electric character. Dielectric probe particles will generally sense the electric part of the electromagnetic momentum and spin densities [2, 7, 8, 13]. Hence, we treat the standard (electric-biased) formulation of the electromagnetic spin and orbital angular momenta. In the presence of dielectric media, such as iron garnets in the near-infrared range, the expression for spin angular momentum becomes

$$\vec{s}_{B,M} = \text{Im} \frac{\varepsilon_o \varepsilon}{2\omega} (\vec{E}^* \times \vec{E}). \tag{Eq. 14}$$

The orbital momentum is

$$\vec{p}_O = \text{Im} \frac{\varepsilon}{2\omega} (\varepsilon_o \vec{E}^* \cdot (\nabla) \vec{E}), \text{ where} \tag{Eq. 15}$$

$\vec{X} \cdot (\nabla) \vec{Y} = X_x \nabla Y_x + X_y \nabla Y_y + X_z \nabla Y_z$, and $\varepsilon$ is the relative dielectric permittivity of the medium [4, a, D, E].

In magneto-optic media, the dielectric permittivity $\varepsilon$ is $\varepsilon_c \pm g$, depending on the helicity of the propagating transverse circular polarization. This is usually a small correction to $\varepsilon_c$, as $g$ is two-, or three-, orders of magnitude smaller in iron garnets, in the near infrared range. For elliptical spins, where one helicity component dominates, we account for the admixture level of the minority component in $\varepsilon$ through a weighted average.

NONRECIPROCAL ELECTROMAGNETIC TRANSVERSE SPIN ANGULAR MOMENTUM AND SPIN-ORBIT COUPLING

1. Transverse Spin Momentum and Angular Momentum Densities in Non-Reciprocal Media

In this section we present a formulation for the transverse-spin momentum and angular momentum densities, as well as the orbital angular momentum density, induced by evanescent fields in nonreciprocal magneto-optic media. The magnitude and tuning range of these objects in terms of waveguide geometry and optical gyrotropy are expounded and discussed. We detail their unequal response to given optical energy fluxes in opposite propagation directions and to changes in applied magnetic fields. And we apply the recently proposed Bliokh-Dressel-Nori electromagnetic spin-orbit correction term to calculate the spin-orbit interaction for evanescent waves in gyrotropic media [7].

Equation (13), together with Eq. (14) and Eq. (15) yield the following expressions for the transverse Belinfante-Minkowski spin angular momentum, spin momentum and the orbital momentum densities in evanescent nonreciprocal electromagnetic waves,

$$\vec{s}_{B,M} = \frac{\varepsilon}{\omega^3 \varepsilon_o} \left( \frac{\varepsilon_c \gamma_{eff} - \beta g}{\varepsilon_c^2 - g^2} \right) \left( \frac{\beta \varepsilon_c - g \gamma_{eff}}{\varepsilon_c^2 - g^2} \right) |H_y|^2 \hat{y} \qquad \text{(Eq. 16)}$$

$$\vec{p}_{B,M} = -\frac{\varepsilon \gamma_{eff}}{\omega^3 \varepsilon_o} \left( \frac{\varepsilon_c \gamma_{eff} - \beta g}{\varepsilon_c^2 - g^2} \right) \left( \frac{\beta \varepsilon_c - g \gamma_{eff}}{\varepsilon_c^2 - g^2} \right) |H_y|^2 \hat{z} \qquad \text{(Eq. 17)}$$

$$\vec{p}_O = \left( \frac{\beta \varepsilon}{2 \omega^3 \varepsilon_o} \left[ \left( \frac{\varepsilon_c \gamma_{eff} - \beta g}{\varepsilon_c^2 - g^2} \right)^2 + \left( \frac{\beta \varepsilon_c - g \gamma_{eff}}{\varepsilon_c^2 - g^2} \right)^2 \right] \right) |H_y|^2 \hat{z} \qquad \text{(Eq.18)}$$

And the ratio

$$\left| \frac{\vec{p}_O}{\vec{s}_{B,M}} \right| = \frac{\beta}{2} \left( \frac{\varepsilon_c \gamma_{eff} - \beta g}{\beta \varepsilon_c - g \gamma_{eff}} + \frac{\beta \varepsilon_c - g \gamma_{eff}}{\varepsilon_c \gamma_{eff} - \beta g} \right) \qquad \text{(Eq. 19)}$$

These expressions depend on the magneto-optic gyrotropy parameter *g* and the dielectric permittivity of the waveguide core channel and of its cover layer under transverse magnetization. They yield different values under magnetic field tuning, magnetization and beam propagation direction reversals, and as a function of waveguide core thickness as discussed below. The propagation constant $\beta$ is gyrotropy-, propagation-direction-, and waveguide-core-thickness-dependent, and this behavior strongly impacts the electromagnetic spin and orbital momenta.

The time-averaged electromagnetic energy flux (Poynting's vector) in the iron garnet layer is

$$\vec{S} = \tfrac{1}{2} \text{Re}(\vec{E}^* \times \vec{H}) = \frac{1}{2} \frac{\beta \varepsilon_c - g \gamma_{eff}}{\omega \varepsilon_0 (\varepsilon_c^2 - g^2)} |H_y|^2 \hat{z}. \qquad \text{(Eq. 20)}$$

Re-expressing the transverse Belinfante-Minkowski spin angular momentum and spin momentum densities in terms of the energy flow $\vec{S}$,

$$\vec{s}_{B,M} = \frac{2\varepsilon}{\omega^2} \left( \frac{\varepsilon_c \gamma_{eff} - \beta g}{\varepsilon_c^2 - g^2} \right) |\vec{S}| \hat{y} \qquad \text{(Eq. 21)}$$

$$\vec{p}_{B,M} = -\frac{2\varepsilon \gamma_{eff}}{\omega^2} \left( \frac{\varepsilon_c \gamma_{eff} - \beta g}{\varepsilon_c^2 - g^2} \right) \vec{S} \qquad \text{(Eq. 22)}$$

Figure 1 plots the nonreciprocal Belinfante-Minkowski transverse spin-angular-momentum-density shift per unit energy flux, as a function of silicon slab thickness in an SOI slab waveguide with Ce$_1$Y$_2$Fe$_5$O$_{12}$ garnet top cladding. Calculations are performed for the same electromagnetic energy flux in opposite propagation directions, at a wavelength of 1550 nm, $g = -0.0086$. The nonreciprocal shift is normalized to the average spin angular momentum, as follows,

$$\Delta \vec{s}_{B,M,S} = \frac{2\left[\varepsilon_f \left(\varepsilon_c \gamma_{eff} - \beta g\right)_f - \varepsilon_b \left(\varepsilon_c \gamma_{eff} - \beta g\right)_b\right]}{\varepsilon_f \left(\varepsilon_c \gamma_{eff} - \beta g\right)_f + \varepsilon_b \left(\varepsilon_c \gamma_{eff} - \beta g\right)_b}. \quad \text{(Eq. 23)}$$

Subscripts *f* and *b* stand for forward, and backward propagation, respectively. This expression evinces a relatively stable value, close to 0.7% above 0.3μm thickness. What is the explanation for this? It has to do with the ellipticity of the transverse polarization in the x-z plane. Above 0.3μm, the ellipticity ranges from 31.4° to 36.9°, where 45° corresponds to circular polarization. In other words, the ellipticity stays fairly constants, with a moderately small admixture of the minority circularly polarized component, ranging from 25% to 14%. Below 0.3μm, the minority component admixture increases precipitously, reaching 87% at 0.13μm.

Magnetization reversals produce the same effect. Consider the nonreciprocal Belinfante-Minkowski transverse spin-angular-momentum-density shift, as a function of silicon slab thickness. Figure 2 plots the normalized shift in Eq. 16 pre-factor,

$$\Delta s_{B,M} = \frac{\left[\varepsilon\left(\frac{\varepsilon_c \gamma_{eff} - \beta g}{\varepsilon_c^2 - g^2}\right)\left(\frac{\beta \varepsilon_c - g \gamma_{eff}}{\varepsilon_c^2 - g^2}\right)\right]_g - \left[\varepsilon\left(\frac{\varepsilon_c \gamma_{eff} - \beta g}{\varepsilon_c^2 - g^2}\right)\left(\frac{\beta \varepsilon_c - g \gamma_{eff}}{\varepsilon_c^2 - g^2}\right)\right]_{-g}}{\frac{1}{2}\left\{\left[\varepsilon\left(\frac{\varepsilon_c \gamma_{eff} - \beta g}{\varepsilon_c^2 - g^2}\right)\left(\frac{\beta \varepsilon_c - g \gamma_{eff}}{\varepsilon_c^2 - g^2}\right)\right]_g + \left[\varepsilon\left(\frac{\varepsilon_c \gamma_{eff} - \beta g}{\varepsilon_c^2 - g^2}\right)\left(\frac{\beta \varepsilon_c - g \gamma_{eff}}{\varepsilon_c^2 - g^2}\right)\right]_{-g}\right\}} \quad \text{(Eq. 24)}$$

We observe the same qualitative thickness dependence as in Fig. 1, corresponding to the moderate, and relatively stable, admixture of minority circularly-polarized component above 0.3μm thickness.

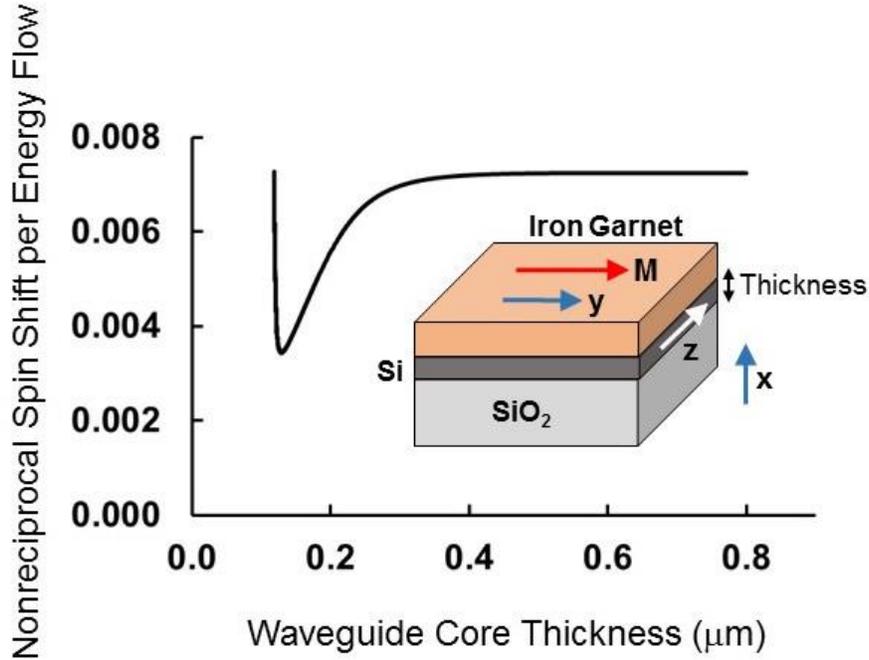

**Fig. 1.** Normalized nonreciprocal Belinfante-Minkowski transverse spin-angular-momentum-density shift per unit energy flux as a function of silicon slab thickness for $g = -0.0086$, corresponding to $Ce_1Y_2Fe_5O_{12}$ garnet top cladding on SOI at $\lambda = 1.55 \mu m$ wavelength. The inset shows the slab waveguide structure. **M** stands for the magnetization in the garnet.

The magneto-optic gyrotropy of an iron garnet can be controlled through an applied magnetic field. These ferrimagnetic materials evince a hysteretic response, such as the one displayed in Fig. 3 (inset) for 532nm wavelength in a sputter-deposited film. The target composition is $Bi_{1.5}Y_{1.5}Fe_{5.0}O_{12}$. Shown here are actual experimental data extracted from Faraday rotation measurements. Below saturation, the magneto-optic response exhibits an effective gyrotropy value that can be tuned through the application of a magnetic field. These measurements correspond to a 0.5μm-thick film on a (100)-oriented terbium gallium garnet (TbGG) substrate. The optical beam is incident normal to the surface, and the hysteresis loop probes the degree of magnetization normal to the surface as a function of applied magnetic field. These data show that the electromagnetic spin angular momentum can be tuned below saturation and between opposite magnetization directions.

Figure 3 also reveals an interesting feature about the magneto-optic gyrotropy. The normalized nonreciprocal Belinfante-Minkowski transverse spin-angular-momentum-density shift per unit energy flux, $\Delta \vec{s}_{B,M,S}$, linearly tracks the gyrotropy, and is of the same order of magnitude as $g$, although thickness-dependent. Yet, as pointed out before, this thickness dependence reflects the admixture of the minor helicity component in the spin ellipticity. At 0.4μm, for example, $\Delta \vec{s}_{B,M,S} = 0.0072$ when $g = -0.0086$. However, the major polarization helicity component contribution to $\Delta \vec{s}_{B,M,S}$ is 84.4% at this thickness, translating into 0.00853 at 100%. At 0.25μm, $\Delta \vec{s}_{B,M,S} = 0.00655$, and the major polarization helicity component contribution is 76.2%, translating into 0.0086 at 100%. *We, thus, re-interpret the magneto-optical gyrotropy as the normalized Belinfante-Minkowski spin-angular-momentum density shift per unit energy flux.*

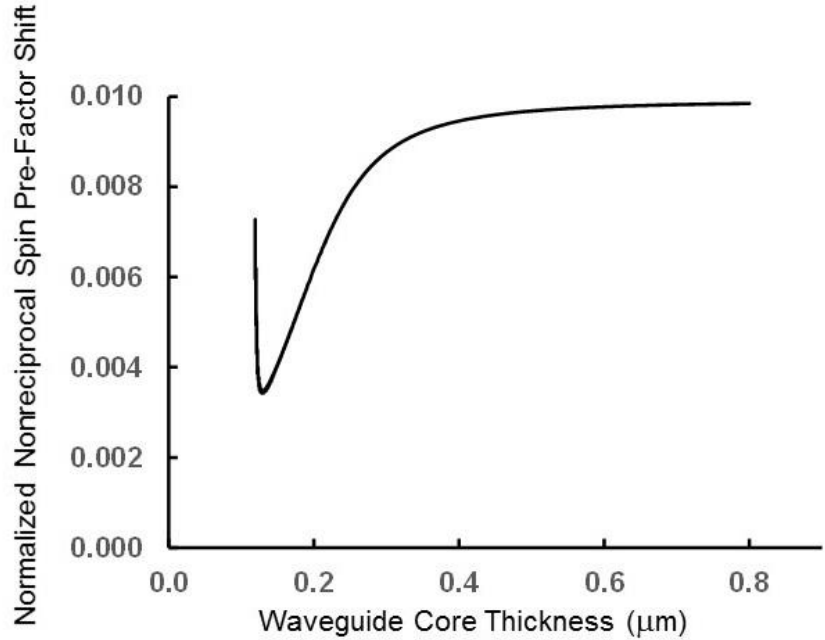

**Fig. 2.** Normalized nonreciprocal Belinfante-Minkowski transverse spin-angular-momentum-density pre-factor shift as a function of silicon slab thickness for $g = -0.0086$, corresponding to $Ce_1Y_2Fe_5O_{12}$ garnet top cladding on SOI at $\lambda = 1.55 \mu m$ wavelength.

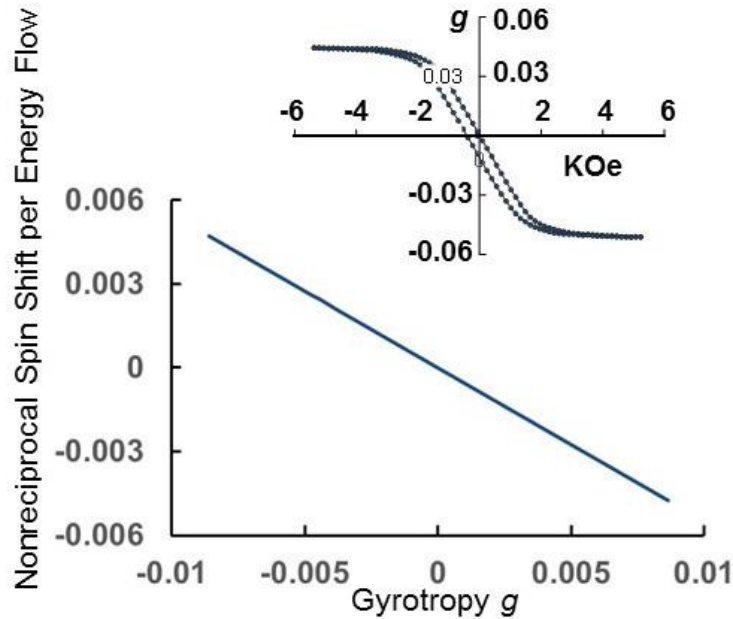

**Fig. 3.** Normalized nonreciprocal Belinfante-Minkowski transverse spin-angular-momentum-density shift per unit energy flux as a function of magneto-optical gyrotropy. Data correspond to 0.25μm silicon-slab thickness with Ce$_1$Y$_2$Fe$_5$O$_{12}$ garnet top cladding, $\lambda = 1.55 \mu m$ wavelength. The inset shows the gyrotropy versus magnetic field hysteresis loop of a magnetic garnet film at $\lambda = 532 nm$, sputter-deposited using a Bi$_{1.5}$Y$_{1.5}$Fe$_{5.0}$O$_{12}$ target.

## 2. Magnetization-Induced Electromagnetic Spin-Orbit Coupling

Bliokh and co-authors have studied the electromagnetic spin-orbit coupling in non-paraxial optical vortex beams [7, 13]. They find that there is a spin dependent term in the orbital angular momentum expression that leads to spin-to-orbit angular momentum conversion. This phenomenon occurs under tight focusing or the scattering of light [7, 13]. Here we consider an alternative source of electromagnetic spin-orbit coupling, magnetization-induced coupling in evanescent waves.

The time-averaged spin- and orbital-angular momenta conservation laws put forth in [7] are

$$\partial_t \operatorname{Im} \frac{1}{2\omega}(\vec{E}^* \times \vec{E})_i = -\nabla_j \operatorname{Im} \frac{\mu_o}{2\omega}\left[\delta_{ij}\vec{H}^* \cdot \vec{E} - H_i^* E_j - H_j^* E_i\right], \text{ and} \qquad (\text{Eq. 25})$$

$$\partial_t \operatorname{Im} \frac{1}{2\omega}\left[\vec{E}^* \cdot (\vec{r} \times \nabla)\vec{E}\right]_i = -\nabla_j \left\{ \operatorname{Im} \frac{\mu_o}{2\omega}\left[\varepsilon_{jkl} H_l^* (\vec{r} \times \nabla)_i E_k + H_j^* E_i\right] + \frac{1}{4}\varepsilon_{ijk}\left(|\mu_o \vec{H}|^2 - |\vec{E}|^2\right)\right\}$$

(Eq. 26)

Latin indices *i, j,…* take on values x, y, z and $\varepsilon_{ijk}$ is the Levi-Civita symbol. Summation over repeated indices is assumed.

The interesting term in these equations, responsible for spin-orbit coupling, is $\operatorname{Im} \frac{\mu_o}{2\omega} H_j^* E_i$. Notice that it appears with opposite signs in the above equations, signaling a transfer of angular momentum from spin to orbital motion. As it stands, so far in our treatment, this term equals zero, since the spin points in the y-direction and the electric-field components of the TM wave point in the x-, and z-directions. A way to overcome this null coupling, and enable the angular momentum transfer, is to partially rotate the applied magnetic field about the x-axis away from the y-direction, as in Fig. 4. This action induces a Faraday rotation about the z-axis, generating a spin-orbit coupling term in the angular momentum conservation laws. A slight rotation or directional gradient in the magnetization **M** will induce electromagnetic spin-orbit interaction in the magneto-optic medium.

Maxwell-Ampere's law acquires off-diagonal components $\pm i\delta g$ in the dielectric permittivity tensor upon rotation of the magnetic moment in the iron garnet film away from the y-axis, as shown in Eq. 27 [14]. Hence, non-zero electromagnetic field components $E_y$ and $H_x$, and spin-orbit coupling, are induced in the propagating wave. The spatial, non-intrinsic, component, characteristic of orbital motion, emerges in the form of a z-dependence in the angular momentum, embodied in the partial or total evanescence of the major circularly-polarized component as the wave propagates along the guide.

$$\vec{\nabla} \times \vec{H} = \varepsilon_o \hat{\varepsilon} \frac{\partial \vec{E}}{\partial t} = -\varepsilon_o \begin{pmatrix} \varepsilon_c & i\delta g & ig \\ -i\delta g & \varepsilon_c & 0 \\ -ig & 0 & \varepsilon_c \end{pmatrix} i\omega \vec{E} \qquad \text{(Eq. 27)}$$

In what sense is there an angular momentum transfer from spin to orbital, in this case? As the polarization rotates in the x-y plane due to the Faraday Effect, there will be a spatially-dependent reduction in the circulating electric field component of the electromagnetic wave along the propagation-direction. This can be seen as a negative increase in circular polarization with z, i.e., an orbital angular momentum in the opposite direction to the electromagnetic spin.

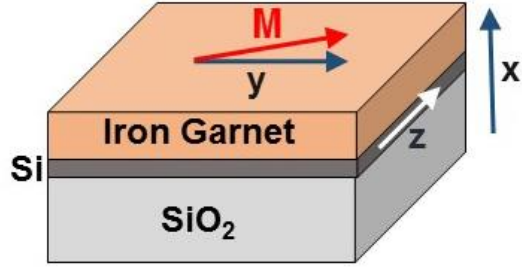

**Fig. 4.** Rotated magnetization **M** generates TM to TE waveguide mode coupling and electromagnetic spin-orbit coupling.

Finally, we derive an explicit expression for the spin-orbit coupling term. The relevant term appearing in the orbital angular momentum flux in the z-direction is

$$-\text{Im} \frac{\mu_0}{2\omega} \partial_z \left[ H_z^* E_y \right] \qquad \text{Eq. (28)}$$

We assume that Faraday rotation induces the $E_y, H_z$ terms via TM to transverse-electric (TE) mode conversion, where

$$H_z = \frac{-i}{\mu_0 \omega} \frac{\partial}{\partial x} E_y, \qquad \text{Eq. (29)}$$

and

$$E_y = E_{y,0} e^{-\gamma_{TE} x} e^{i\beta_{TE} z} \sin(\theta_F z). \qquad \text{Eq. (30)}$$

$E_{y,0}$ is the electric field amplitude corresponding to full TM to TE conversion, $\theta_F$ is the specific Faraday rotation angle, $\gamma_{TE}$ and $\beta_{TE}$ are the cover-layer decay constant and the propagation constant for the TE mode, respectively. For simplicity, we assume no linear birefringence in the waveguide, so $\beta_{TE} = \beta_{TM}$. Hence, the spin to orbital angular momentum coupling term is

$$-\text{Im} \frac{\mu_0}{2\omega} \partial_z \left( H_z^* E_y \right) = \frac{\gamma_{TE}}{2\omega^2} \left| E_{y,0} \right|^2 e^{-2\gamma_{TE} x} \theta_F \sin(2\theta_F z) \qquad \text{Eq. (31)}$$

Acknowledgment

The authors thank Ramy El-Ganainy for suggesting this problem and for useful discussions.